\documentclass[10pt,preprint]{aastex}

\usepackage{afterpage}
\usepackage{pifont}
\usepackage{aas_macros}
\usepackage{natbib}
\usepackage{footnote}
\usepackage{threeparttablex,longtable,multirow} 
\usepackage{moresize}
\usepackage{color,colortbl}
\usepackage{epstopdf}

\usepackage{geometry} 
\usepackage{graphicx} 
\usepackage{booktabs} 
\usepackage{array} 
\usepackage{paralist} 
\usepackage[caption=false]{subfig} 
\usepackage{amsmath,amssymb}

\usepackage{fancyhdr} 
\begin{document}
\title{Exoplanet Predictions Based on the Generalised Titius-Bode Relation}

\author{Timothy Bovaird\altaffilmark{1,2}, Charles H. Lineweaver\altaffilmark{1,2,3}}
\altaffiltext{1}{Research School of Astronomy and Astrophysics, Australian National University, Canberra, Australia}
\altaffiltext{2}{Planetary Science Institute, Australian National University}
\altaffiltext{3}{Research School of Earth Sciences, Australian National University}


\begin{abstract}
We evaluate the extent to which newly detected exoplanetary systems containing at least four planets adhere to a generalized Titius-Bode (TB) relation. We find that the majority of exoplanet systems in our sample adhere to the TB relation to a greater extent than the Solar System does, particularly those detected by the Kepler mission.
We use a generalized TB relation to make a list of predictions for the existence of 141 additional exoplanets in 68 multiple-exoplanet systems: 73 candidates from interpolation, 68 candidates from extrapolation. 
We predict the existence of a low-radius ($R<2.5R_\oplus$) exoplanet within the habitable zone of KOI-812 and that the average number of planets in the habitable zone of a star is 1-2.
The usefulness of the TB relation and its validation as a tool for predicting planets will be partially tested by upcoming Kepler data releases.
\end{abstract}
\maketitle

\section{INTRODUCTION}
\label{sec:intro}
During the last few years the number of multiple-exoplanet systems has increased rapidly. While most multiple-exoplanet systems contain two or three planets, a rapidly growing number of systems now contain at least four planets. The architecture of these systems can begin to be compared to our Solar System. Many exoplanet systems appear to be much more compact and their planets more evenly spaced than the planets of our Solar System. This paper sets out to quantify these tendencies and place constraints on the location of undetected planets in multiple-exoplanet systems. 

The approximately even logarithmic spacing between the planets of our Solar System motivated the Titius-Bode (TB) relation, which played an important role in the discovery of the Asteroid Belt and Neptune, although Neptune was not as accurately represented by the TB relation as the other planets \citep{SawyerHogg1948,Lyttleton1960,Brookes1970,Nieto1972}. Uranus could have been discovered earlier if the TB relation had been take more seriously \citep{Nieto1972}. Since the TB relation was a useful guide in the prediction of undetected planets in our Solar System, it may be useful for making predictions about the periods and positions of exoplanets in multiple-exoplanet systems. The main goal of this paper is to test this idea.
We first examine the degree to which multiple-exoplanet systems adhere to the TB relation. Due to limited detection sensitivity, we have no reason to believe that all exoplanets in a given system have been detected.
After assessing this incompleteness in several ways, we find that the more complete the exoplanet system, the more it adheres to the TB relation. If this trend is correct, the TB relation can be profitably applied to multi-planet systems to predict the periods, positions and upper mass or radius limits for as yet undetected exoplanets.

The distribution and separation of planets in our Solar System can be parameterized by a generalized (two parameter) TB relation \citep{Goldreich1965,Dermott1968}:
\begin{equation}
a_n=a C^n, \hspace{10pt}n=0,1,2,3,...
\label{eq:tbpowerlaw}
\end{equation}
where $a_n$ is the semi-major axis of the $n$th planet or satellite, $a$ is the best-fit parameter associated with the semi-major axis $a_0$ of the first planet, 
and the powers of $C$ parametrize the logarithmic spacing. This form of the TB relation involves two free parameters $(a\text{ and }C)$, instead of three from the original TB relation (\citet{Wurm1787}, Equation~\ref{eq:appendix3TB}), which was derived emperically based on our Solar System only. 

The paper is organized as follows. Section~\ref{sec:physics} discusses the physics behind the TB relation. In Section~\ref{sec:previouswork} we review recent work on the TB relation and planet predictions that have been made based on it. In Section~\ref{sec:method} we describe our analysis and present our main results (predictions for exoplanet periods). We discuss the details of several systems, caveats and compare our results with previous studies in Section~\ref{sec:discussion}. We summarize our results in Section~\ref{sec:conclusion}. 

\section{PHYSICS BEHIND THE TB RELATION}
\label{sec:physics}
As proto-planetary oligarchs accrete mass through collisions within a protoplanetary disk, they clear out the material in nearby orbits (as in the IAU definition of a planet\footnote{http://www.iau.org/static/resolutions/Resolution\_GA26-5-6.pdf}), thereby excluding the presence of nearby planets.
Starting from a relatively smooth disk (e.g. a surface density $\Sigma\propto r^{-3/2}$), the areas of mutual exclusion grow with time as the system relaxes dynamically. This produces a non-random distribution of planets in which the planet periods and positions are roughly logarithmically spaced (at least in the case of our Solar System, \citet{Hayes1998}). 

A roughly logarithmic spacing between planet periods can be roughly parametrized by the generalized TB relation. 
Due to the stochastic nature of planetary scattering, planet migration, high eccentricities, and other variables of planetary formation, it is not obvious that the TB relation should even approximately fit exoplanet systems. 

From Equation~\ref{eq:tbpowerlaw} and Kepler's 3rd law ($P_n \propto a_n^{3/2}$), the generalized TB relation can be restated in terms of the periods of the orbits,
\begin{align}
P_n=P \alpha^n, \hspace{10pt}n=0,1,2,...N-1
\label{eq:perTBpowerlaw}
\end{align}
where $P_n$ is the period of the $n$th planet or satellite, and $P$ and $\alpha (= C^{3/2})$ are parameters to be fit for each system. 

It is well known that the period ratios of planets and satellites in the Solar System show a preference towards near mean motion resonance (NMMRs) \citep{Roy1954,Goldreich1965,Dermott1968a}. More recently it has been shown that planets in exosystems also exhibit the same preference towards near mean motion resonances \citep{Batygin2012,Lithwick2012,Fabrycky2012}. From Equation~\ref{eq:perTBpowerlaw} the period ratio between adjacent orbits is $P_{n+1}/P_n=\alpha$. If $\alpha$ is approximately constant for all $n$ (all planet pairs in the system), then the generalized TB relation will fit the system well. The physical mechanisms underlying these empirical observations are not fully understood.

For example, Jupiter and Saturn are in a $5:2$ NMMR. Approximately the same NMMR exists between Mercury and Venus, Mars and the Asteroid Belt and the Asteroid Belt and Jupiter. The remaining planet pairs have period ratios of $\sim1.6$ ($\sim3:2$), $\sim1.9$ ($\sim2:1$), $\sim2.9$ ($\sim3:1$) and $\sim2.0$ ($\sim2:1$) for Venus and Earth, Earth and Mars, Saturn and Uranus and Uranus and Neptune respectively. 
The larger the scatter around the dominant NMMR, the less well the system adheres to the generalized TB relation. In the case of our Solar System, the smaller the scatter of period ratios around the dominant $5:2$ ratio, the better the fit to the generalized TB relation.

\section{PREVIOUS WORK USING THE TB RELATION}
\label{sec:previouswork}

\cite{Hills1970} analyzed adjacent planet period ratios in 11 simulated systems. Each system was evolved with a 1 $M_\odot$ host star but with different planet mass distributions. He suggested that the TB relation results from 
some NMMRs ($5:2$ and $3:2$) being more stable than others.
 Instability tends to exclude certain period ratios between interacting planets. The stronger the interactions, the more exclusion in period ratio space, the narrower the bunching of periods ratios, and the better the fit to the TB relation. \cite{Laskar2000} performed simplified numerical simulations of $10^5$ planetesimals distributed in systems with different initial planetesimal surface densities $\Sigma(a)$, where $a$ is the semi-major axis. A merger resulted when two planetesimals underwent a close encounter and simulations continued until no more close encounters were possible. The final location of planets could be well fit by a TB relation when the surface density $\Sigma(a)\propto a^{-3/2}$. This same form of $\Sigma$ is derived for the Solar System from the Minimum Mass Solar Nebula \citep{Weidenschilling1977,Hayashi1981} and a similar form, $\Sigma(a)\propto a^{-1.6}$ has been recently derived for a Minimum Mass Extrasolar Nebula from Kepler data \citep{Chiang2012}. 

\cite{Isaacman1977} created model planetary systems by injecting accretion nuclei into a disk of gas and dust until the dust was depleted. They found that simulated systems adhered to the generalized TB relation ``about the same as the Solar System." Another numerical approach was taken by \cite{Hayes1998} where model planetary systems were generated and subjected to simple stability criteria. For example, systems whose adjacent planets were separated by less than $k$ times the sum of their Hill radii, where $k$ ranged from 0 to 8, were discarded. They found that the adherence to the TB relation increases with $k$, i.e. as the region of semi-major axis space open for adjacent planets is restricted, the adherence to the TB relation increases. In period space, as the effect of mutual exclusion increases, the NMMRs closer to 1 are excluded and this results in a narrowing of the spread of period ratios in the system. 

\cite{Poveda2008} looked for the possibility of additional planets in the five-planet 55 Cnc system based on the generalized TB relation (Equation~\ref{eq:tbpowerlaw}).
They predicted planets at $\sim2$ AU and at $\sim15$ AU but made no attempt to quantify the uncertainties of those predictions. \cite{Lovis2011} applied the generalized TB relation to multiple-exoplanet systems observed with HARPS and found reasonable fits, although they made no planet predictions. \cite{Cuntz2011} applied the three-parameter TB relation (Equation~\ref{eq:appendix3TB}) to 55 Cnc. He proposed that should the TB relation be applicable to the sytem, undetected planets may exist at semi-major axes of 0.085 AU, 0.41 AU, 1.50 AU and 2.95 AU. Additionally, a planet exterior to the outermost detected planet was predicted to exist between 10.9 and 12.6 AU. For a comparison of the two and three parameter TB relations see Section~\ref{sec:2v3param} and Appendix~\ref{sec:2v3paramapp}.

\cite{Chang2008} analyzed the period ratios of pairs of planets in individual multiple-planet systems, comparing the distributions of the Solar System and 55 Cnc to that expected from a generalized TB relation (Equation~\ref{eq:tbpowerlaw}). \cite{Chang2010}, repeated the analysis for 31 multiple-exoplanet systems.
Even without taking into account any correction for the expected incompleteness of planet detections, \cite{Chang2010} tentatively concluded that the adherence of the planets of our Solar System to a TB relation is not ``due to chance'' and that one cannot rule out the possibility that the TB relation can be applied to exosystems.

By utilizing numerical N-body simulations one can estimate the regions of semi-major axis and eccentricity space which could host an additional planet without the system becoming dynamically unstable.
Although planets do not necessarily have to occupy these stable regions, \cite{Barnes2004} argue that most planetary systems lie near
instability, leading to the `packed planetary systems' (PPS) hypothesis \citep{Raymond2005a,Raymond2006,Raymond2008,Raymond2009a}. 
The PPS hypothesis suggests that the majority of systems will be near the edge of dynamical instability. That is, undetected planets will occupy a stable region and will tend to bring the system closer to (but not over the edge of) instability. 

The PPS hypothesis and the TB relation are not orthogonal conceptions. As the spacing between planets decreases, i.e. as a system becomes more tightly packed, a system's adherence to the TB relation increases (see Section~\ref{sec:compactness}). The PPS hypothesis suggests that planets are spaced as tightly as possible without the system becoming dynamically unstable. Since dynamical stability is correlated with planets being evenly spaced, we may then expect systems which adhere to the PPS hypothesis to be more likely to adhere to the TB relation. However, the degree to which PPS systems adhere to the TB relation is poorly understood since PPS analyses predominately focus on two-planet systems \citep{Barnes2004,Fang2012}. 

While numerical simulations can be useful for predicting stable regions, computing constraints typically confine direct orbit integrations to two-planet systems. It becomes inefficient to integrate a large number of systems, each containing a large number of planets. There are now many systems containing four or more planets, and this number is growing rapidly due to the Kepler mission. The remainder of this paper discusses our tool for analyzing a large number of systems and constraining regions where we expect undetected planets to exist based on our generalized TB relation. 

\section{METHOD AND RESULTS}
\label{sec:method}
Here we improve on previous TB relation investigations first by more carefully normalizing the $\chi^2/\text{dof}$ fit of the TB relation to our Solar System 
and then by fitting the TB relation to exoplanet systems which contain at least four planets (Figure~\ref{fig:period_systems}). We choose this minimum number of planets as a compromise between, on the one hand, having enough systems to analyze without suffering too badly from small number statistics, and on the other hand, having enough planets in each system to fit 2 parameters and minimize incompleteness of the systems. 

As of 17 July 2013, online exoplanet archives\footnote{NASA Exoplanet Archive (http://exoplanets.org/), Extrasolar
Planets Encyclopaedia (http://exoplanet.eu/), Exoplanet Orbit Database (http://exoplanets.org/) contained 71 systems with at least four planets. We exclude two of these systems, KOI-2248 and KOI-284, based on concerns of false positives in those systems \citep{Lissauer2011a,Fabrycky2012}. We also exclude KOI-3158 due to the current lack of reported stellar parameters. The remaining 68 systems are our sample referred to in the remainder of this paper.}
Seven of the 68 systems were detected by the radial velocity method; Gl 876, mu Ara, ups And, 55 Cnc, Gl 581, HD 40307 and HD 10180. One system, HR 8799, was detected by direct imaging. The remaining 60 are candidate systems detected by the Kepler mission. Four of the Kepler systems have been confirmed by radial velocity observations (Kepler-11, Kepler-20, Kepler-33 and Kepler-62). The remaining 56 Kepler detected systems consist of planet candidates. However, \cite{Lissauer2012} conclude that $< 2\%$ of Kepler candidates in multiple planet systems are likely to be false positives.

Since the jostling during planetary system formation is a messy process that can include planetary scattering and migration, we expect varying degrees of adherence to the TB relation.
Previous TB relation studies of exoplanetary systems \citep{Poveda2008,Cuntz2011} chose where to propose additional planets based on the solution which minimizes $\chi^2/\text{dof}$. 
If our Solar System is treated like an exosystem, where limited detection sensitivity precludes the detection of some planets, this method does not necessarily recover the complete Solar System. 
We use the Solar System as a guide to how well we expect exosystems to adhere to the TB relation. In log space, the period form of the generalized TB relation (Equation~\ref{eq:perTBpowerlaw}) is given by
\begin{equation}
\log P_n=\log P+n\log\alpha=A+Bn, \hspace{10pt}n=0,1,2,...,N-1
\label{eq:log_perTBpowerlaw}
\end{equation} 
where A and B are parameters being fit to each system. The $\chi^2/\text{dof}$ value for a system containing N planets can then be calculated by
\begin{equation}
\dfrac{\chi^2 (A,B)}{N-2}=\dfrac{1}{N-2} \sum_{n=0}^{N-1} \left( \dfrac{(A+Bn)-\log P_n}{\sigma_n} \right)^2
\label{eq:chisquared}
\end{equation}
where $\sigma_n$ is the unknown uncertainty associated with the various reasons that the log of the period of the $n$th planet does not conform to the TB relation. 
The $\sigma_n$ are not the uncertainties in the precisions of measurements of $\log P_n$ because
the deviations of $\log P_n$ from a good fit to the TB relation are dominated by the uncertainties in the complicated set of variables that determine the degree 
to which the NMMRs of adjacent planets bunch up around the same ratio. This is poorly understood \citep{Hills1970, Hayes1998}. 
In our analysis, we choose uniform uncertainties ($\sigma_n = \sigma$) 
and scale $\sigma$ for each system by the mean log period spacing between planets in each system, which we refer to as the sparseness or compactness of the system 
(see Equation~\ref{eq:compactness} and Figure~\ref{fig:compactness_sigma_TB2}).
$N-2$ is the number of degrees of freedom (N planets - 2 fitted parameters (A and B)).

\begin{figure}[p]
\plotone{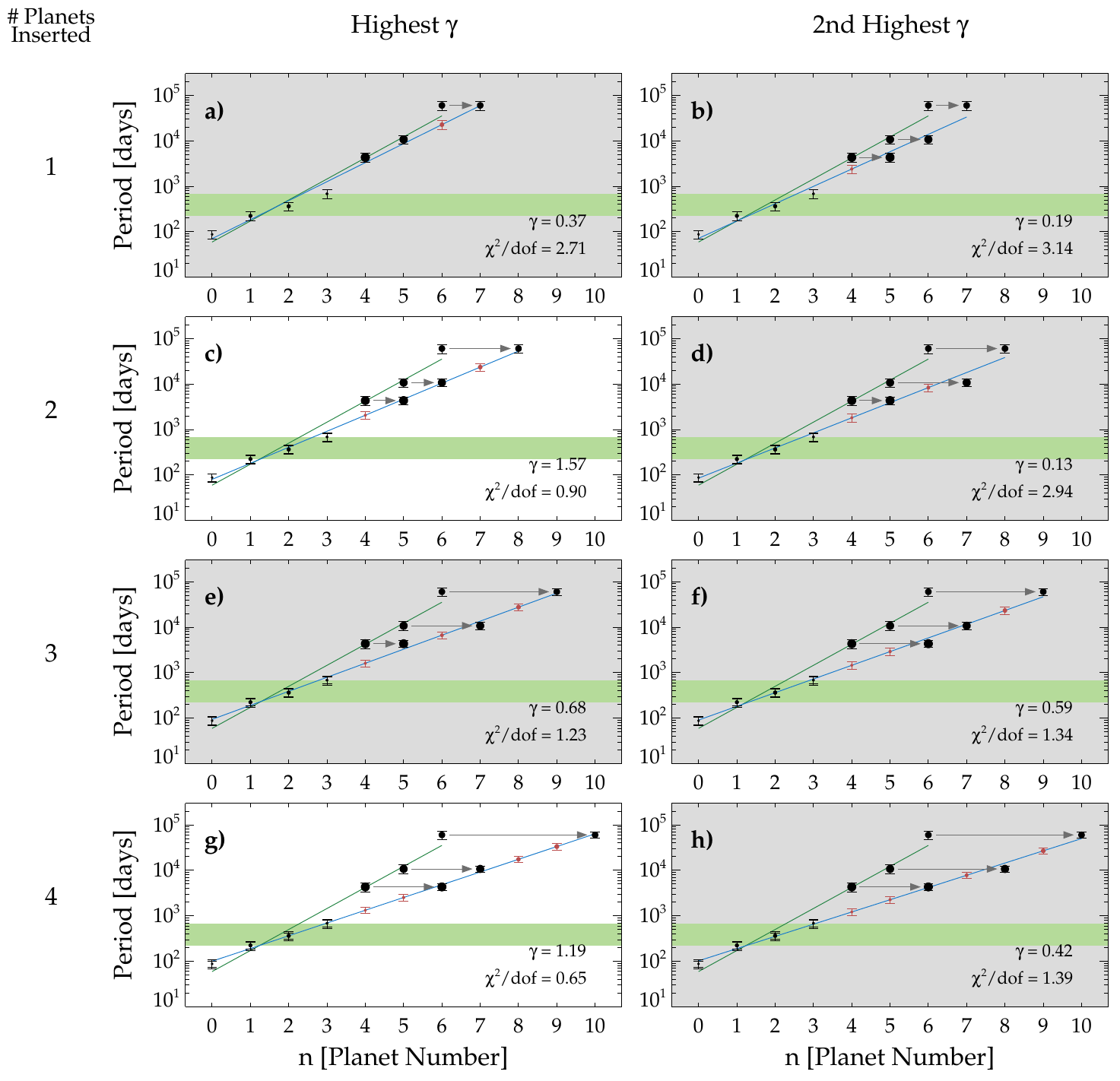}
\caption{\small Fitting the TB relation to the planets of our Solar System when Uranus and the Asteroid Belt are not included. This figure shows the ability of the TB relation to predict where planets could be inserted to reconstruct the incompletely detected planetary system. 
We use Equation~\ref{eq:chisquared} to compute $\chi^2$ with $\text{dof}=7-2=5$. 
We normalize the Solar System's adherence to the TB relation by adjusting $\sigma$ to ensure that $\chi^2/\text{dof}=1$ when Uranus and the Asteroid belt are included in the analysis (dof $=9-2=7$). 
The removal of Uranus and the Asteroid Belt
increases the $\chi^2/\text{dof}$ from 1 to 3.7. 
The green and blue lines show the best fit generalized TB relation (minimizing $\chi^2/\text{dof}$ of Equation~\ref{eq:chisquared}) before and after the insertion of 1 (top row), 2 (second row), 3 (third row) and 4 (fourth row) planets. 
We use the maximum value of $\gamma$ (Eq. 5) to prioritize these fits.
The parameter $\sigma$ has been scaled in each panel according to the sparseness/compactness of the system (Equation 6) as described in Figure 4.
This figure of the Solar System has been constructed to be as comparable as possible to our method of predicting the periods and positions of undetected exoplanets in multi-planet exoplanetary systems. Figure~\ref{fig:55cnc} shows the same method applied to the 55 Cnc system.}
\label{fig:tbsolsys}
\end{figure}

Given the periods of the planets in an exoplanet system that contains 4 planets (e.g. $P_0, P_1, P_2, P_3 $), our strategy is to insert a new fifth planet (or more) between two adjacent planets and then fit the new system (e.g. $P_0, P_1, P_2, P_3, P_4,...$) to the TB relation to see if the new $\chi^2/\text{dof}$ from Equation \ref{eq:chisquared} is lower. The maximum mass or radius of the inserted planet is determined by the minimum signal to noise ratio of detected planets in the same system (Section~\ref{sec:constraints}). An example of this is shown in Figure~\ref{fig:tbsolsys} where we remove the Asteroid Belt and Uranus
(the two Solar System objects 
whose approximate periods
were predicted by the TB relation before their discovery) from the  Solar System and then 
insert planets and compute the $\chi^2/\text{dof}$ fit to the TB relation. When the Asteroid Belt and Uranus are included in the fit to the Solar System, the $\chi^2/\text{dof}$ decreases from $\sim3.7$ to $\sim1.0$, thus improving the fit to the TB relation. 

We do not increase the dof when we insert planets because the new planets are inserted on the best fit line and do not contribute to $\chi^2$. However, inserting planets does increase the compactness and therefore, as shown in Figure 4, does reduce the value of $\sigma$.
In some cases the insertion of an additional planet into a system results in a decrease of $\chi^2/\text{dof}$ which means that the new system with the inserted planet adheres to the TB relation better than the system without the insertion.

Of the many possible ways to insert planets, we want to identify the way which improves the adherence to the TB relation the most. At the same time, we 
want to avoid inserting too many planets, since the lowest $\chi^2/\text{dof}$ does not necessarily correspond to the complete system (at least in the case of the Solar System). See for example,
panels c and g of Figure~\ref{fig:tbsolsys}. 
To quantify this, we define a variable $\gamma$, which is a measure of the fractional amount by which the $\chi^2/\text{dof}$ improves, divided by the number of inserted planets (i.e. the improvement in the $\chi^2/\text{dof}$ per inserted planet):

\begin{equation}
\gamma=\frac{\left(\dfrac{\chi^2_i-\chi^2_f}{\chi^2_f}\right)}{n_\text{ins}}
\label{eq:gamma}
\end{equation}
where $\chi^2_i$ and $\chi^2_f$ are the $\chi^2$ values from Equation~\ref{eq:chisquared} before and after the insertion of $n_\text{ins}$ planets. $\gamma$ is calculated for each insertion solution and the solution with the highest $\gamma$ is chosen. 
Choosing the highest $\gamma$ recovers the actual Solar System when the input system does not include Uranus and the Asteroid Belt (Figure~\ref{fig:tbsolsys}). 
This result is sensitive to which objects are removed from the Solar System. In the special case of an outside observer performing a $\sim 30$ year radial velocity survey of the Solar System with a detection sensitivity of $\sim8$ cm$^\text{-1}$, Venus, Earth, Jupiter and Saturn could be detected. Applying the above method to this system results in the prediction of planets at the orbital periods of Mars and the Asteroid Belt. If we allow the fit to be extended to planets interior to Venus or exterior to Saturn, the remaining Solar System planets are also recovered.

In the more general case (removing all possible combinations of two planets from the Solar System), the highest $\gamma$ value recovers the complete Solar System $38\%$ of the time. Similarly, when removing all possible combinations of three planets, the complete Solar System is recovered $29\%$ of the time. 

\begin{figure}[p]
\epsscale{0.97}
\plotone{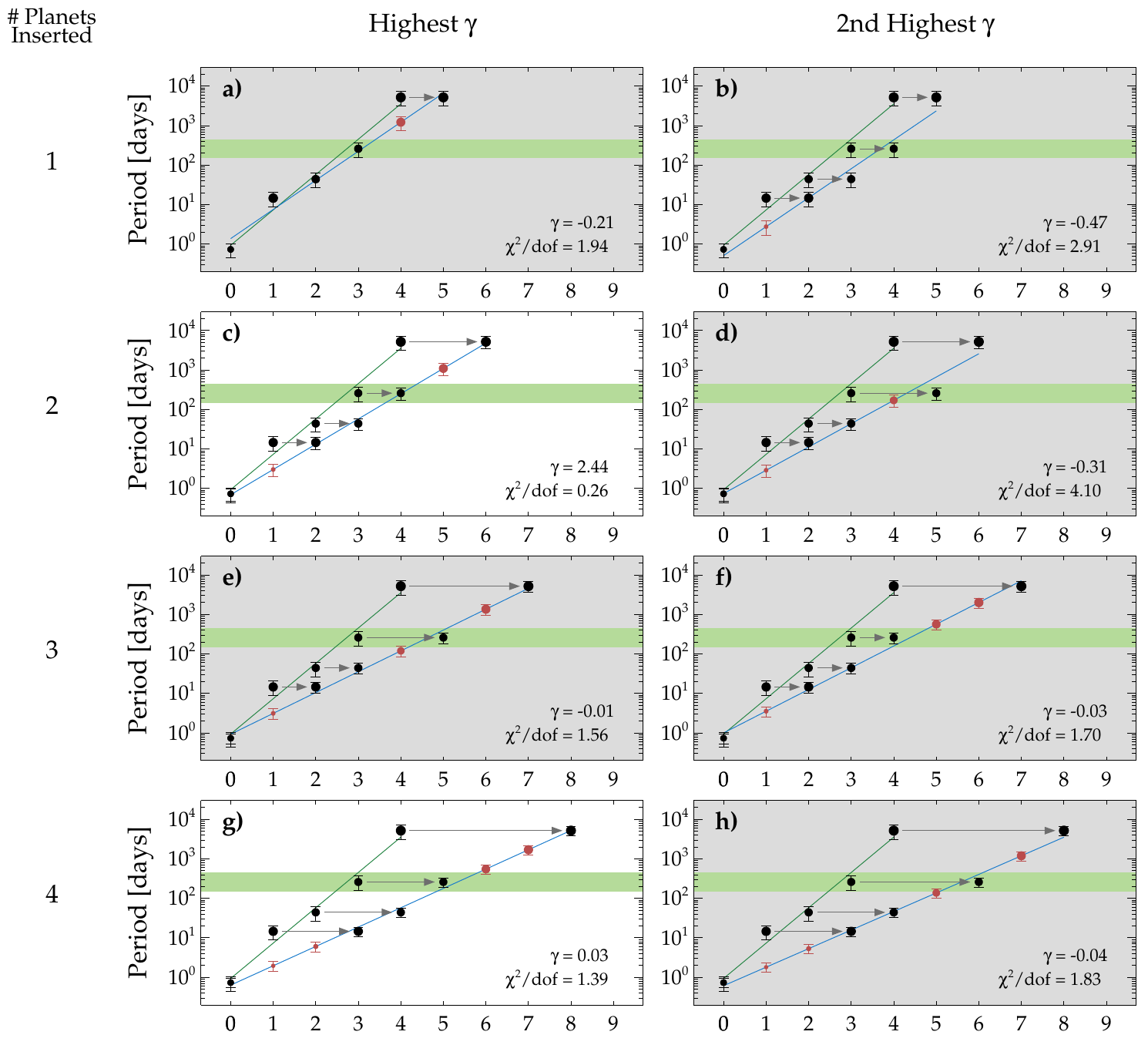}
\epsscale{1.0}
\caption{\small Illustration of our method for the 55 Cnc system, which has an initial $\chi^2/\text{dof}$ of 1.54. One to four planets are inserted between detected planets and new $\sigma$ and $\chi^2/\text{dof}$ values are calculated for each insertion solution. For each number of planets inserted from 1 (top panels) to 4 (bottom panels), the two lowest $\chi^2/\text{dof}$ solutions are displayed. The insertion solutions with the two highest $\gamma$ values (Equation 5) are indicated by a white background. An approximate habitable zone is represented by the horizontal green bar, which represents the effective temperature range between Venus and Mars, using an albedo of 0.3.}
\label{fig:55cnc}
\end{figure}

\definecolor{grey1}{rgb}{0.863,0.863,0.863}
\definecolor{grey2}{rgb}{0.686,0.686,0.686}
\begin{table}[p]
\caption{Data corresponding to Figure~\ref{fig:55cnc}}
\label{table:55cnc}
\centering
\scriptsize
\renewcommand{\arraystretch}{0.5}
\setlength\minrowclearance{2.0pt}
\newcolumntype{C}[1]{>{\centering\arraybackslash}p{#1}}
\begin{tabular}{C{30pt}C{50pt}C{38pt}C{25pt}C{53pt}C{53pt}C{53pt}C{53pt}}
\toprule
\multirow{2}{*}{\bf Panel} & \multirow{2}{*}{\bf \hspace{-5pt}\# Inserted} & \multirow{2}{*}{\bf $\boldsymbol\chi^2/\text{dof}$} & \multirow{2}{*}{\bf $\boldsymbol\gamma$} & {\bf Pl. Period 1} & {\bf Pl. Period 2} & {\bf Pl. Period 3} & {\bf Pl. Period 4} \\
 & & & & {\bf (days)} & {\bf (days)} & {\bf (days)} & {\bf (days)} \\
\midrule
- & 0 & 1.54 & - & - & - & - \\
\rowcolor{grey1}
a & 1 & 1.94 & -0.21 & 1227.7 & - & - & - \\
\rowcolor{grey1}
b & 1 & 2.91 & -0.47 & 2.8 & - & - & - \\
c & 2 & 0.26 & 2.44 & 3.0 & 1103.6 & - & - \\
\rowcolor{grey1}
d & 2 & 4.10 & -0.31 & 2.9 & 171.7 & - & - \\
\rowcolor{grey1}
e & 3 & 1.56 & -0.01 & 3.1 & 120.0 & 1363.9 & - \\
\rowcolor{grey1}
f & 3 & 1.70 & -0.03 & 3.6 & 567.3 & 2015.5 & - \\
g & 4 & 1.39 & 0.03 & 2.0 & 6.1 & 552.3 & 1705.9 \\
\rowcolor{grey1}
h & 4 & 1.83 & -0.04 & 1.8 & 5.3 & 137.7 & 1200.6 \\
\bottomrule
\end{tabular}
\end{table}

\subsection{Isolating The Most Complete Systems}
\label{sec:mostcomplete}
To test the adherence of exoplanetary systems to the TB relation, we want to minimize the effects of incompleteness on the test. Thus we identify a more complete sample using a dynamic spacing criterion.
After identifying a sample of exoplanet systems or subsets of exosystems which are likely to be more complete, we test their adherence to the TB relation. A simple approximation of the stability of a pair of planets is the dynamical spacing $\Delta$ \citep{Gladman1993,Chambers1996}, which is a measure of the separation of two planets in units of their mutual Hill radius (Equation B1, Appendix~\ref{sec:delta}). The dynamic spacing $\Delta$ is a function of the planet masses and hence dependent on our radius-mass conversion for transiting planets ($R\approx1.35M^{0.38}$, derived from 26 $R<8R_\oplus$ exoplanets with mass and radius measurements). 

\begin{figure}[p]
\label{fig:spacing_and_resonance_histo}
\plotone{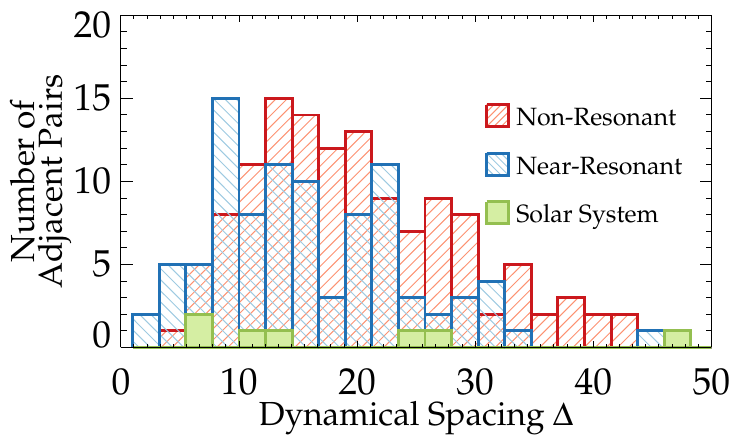}
\caption{\small Dynamical spacing $\Delta$ for adjacent planet pairs within exoplanet systems containing four or more planets, separated into NMMR and non-NMMR planet pairs. The threshold to be in NMMR is set arbitrarily at $x \le 2$\%, where $x = |N_j/N_i-P_{n+1}/P_n|/(N_j/N_i)$. $N_i$ and $N_j$ are positive integers with $N_i < N_j \le 7$ and $P_n$ and $P_{n+1}$ are the periods of the planet pair. There is a drop-off of planet pairs at $\Delta \lesssim 10$ as found in simulations \citep{Chambers1996,Fang2012}. The values for the Solar System are shown for reference.}
\end{figure}

For small eccentricities and small mutual inclinations, \cite{Chambers1996} found the stability of simulated systems increased exponentially with $\Delta$. A disrupting close-encounter between planets was likely within $10^8$ years for $\Delta \lesssim 10$. The same $\Delta \lesssim 10$ stability threshold was observed by \cite{Fang2012} and in a later paper it was shown that the average $\Delta$ value for Kepler detected exoplanet pairs was 21.7 \citep{Fang2013}. The $\Delta$ values for all adjacent pairs of exoplanets in our sample is shown in Figure~\ref{fig:spacing_and_resonance_histo}. There is a drop-off below $\Delta\approx10$, in agreement with the simulations. Also, NMMR pairs dominate the $\Delta\lesssim10$ pairs.
Planet pairs with small dynamic spacing, $\Delta\lesssim10$, are less likely to be stable,
but if they are stable, they are more likely to be NMMR pairs.
We use this criterion to determine which pairs are more likely to be in complete systems where the insertion of an additional planet (with an assumed mass of $1$ M$_\oplus$) between the pair would result in a lower probability of stability: $\Delta\lesssim10$.
Note that this assumed mass for the inserted planet plays a minor role, since the mass of the detected planet is always larger and dominates in the calculation of $\Delta$ (Equation B1, Appendix~\ref{sec:delta}).

If the insertion of an additional planet between each pair in a system results in $\Delta\lesssim10$ for all pairs, then that system is likely to be complete without any insertions. 
Nine systems satisfy this criterion, they are KOI-116, KOI-720, KOI-730, KOI-1278, KOI-1358, KOI-2029, KOI-2038, KOI-2055 and HR 8799. A further 22 systems have 3 or more adjacent planet pairs with $\Delta\lesssim10$ (after insertion) and we treat these subsets (without insertions) as their own complete system, giving a total of 31 systems in our most-complete sample.

Of the 31 systems in our most-complete sample, 26 fit the TB relation better than the Solar System ($\chi^2/\text{dof}<1$, with $\sigma$ scaled with compactness as shown in Figure 4). Three systems fit approximately the same as the Solar System and 2 of the 31 systems fit worse than the Solar System $(\chi^2/\text{dof}\gtrsim1.4)$. If additional planets exist in these systems, they are likely to be in a near mean motion resonance with one or more of the detected planets.

Considering only the 31 systems which are most likely to be complete, $\sim94\%$ ($\approx29/31$) adhere to the TB relation to approximately the same extent or better than the Solar System.
Thus planetary systems, when sufficiently well sampled, have a strong tendency to fit the TB relation. Taking this strong-tendency-to-fit
as a common feature of planetary systems, we make predictions
about the periods and positions of exoplanets that have not yet been detected. Specifically, we fit the TB relation to systems that are less complete, insert planets into a variety of positions and find the positions that maximize the $\gamma$ value. Figure~\ref{fig:tbsolsys} illustrates our procedure for the Solar System when Uranus and the Asteroid Belt are missing.
Figure~\ref {fig:55cnc} and Table~\ref{table:55cnc} illustrate our procedure applied to the 55 Cnc system.
The usefulness of planet predictions based on the TB relation depends on how much of a better fit each insertion solution 
produces (Eq. 5). 

To test the validity of this idea further, we removed individual planets from systems in our most-complete sample, thus creating less-complete systems.
We then inserted a planet in the location predicted by our procedure of maximizing $\gamma$. The predicted location was between the correct pair of adjacent planets $100\%$ of the time.
The predicted period (a new $P_n$ in Equation~\ref{eq:chisquared}) was within our $1\sigma$ of the original period $\sim86\%$ of the time.

\subsection{The Compactness of Planetary Systems}
\label{sec:compactness}
One factor not considered by other studies is the relation between the sparseness or compactness of a system and its adherence to the TB relation.
Our initial analyzes of exoplanet systems used the value of $\sigma$ that made $\chi^2/\text{dof} = 1$ for the Solar System. However, this led to $\chi^2/\text{dof} < < 1,$
particularly for the most complete (and compact) exoplanetary systems.
The Solar System is much more sparse than these systems (Figure 5).
We have included the Solar System in Figure 5 to show this difference. 
In more compact systems with shorter periods, one expects the tendency towards commensurability of neighboring orbits to propagate more easily to other orbits \citep{Goldreich1965,Dermott1968a}. This is especially important when analyzing systems discovered by the Kepler mission, where individual systems have multiple planets within $\sim50$ day periods ($\sim0.3$ AU). 
Thus, we introduce a scaling for $\sigma$ dependent on how compact or sparse the system is.
We define the sparseness $S_p$ of a system as,

\begin{equation}
S_p=\dfrac{\log P_{N-1}-\log P_0}{N}
\label{eq:compactness}
\end{equation}

where $P_{N-1}$ and $P_0$ are the largest and smallest planet periods in the system respectively and $N$ is the number of planets in the system.

\begin{figure}[p]
\epsscale{0.65}
\plotone{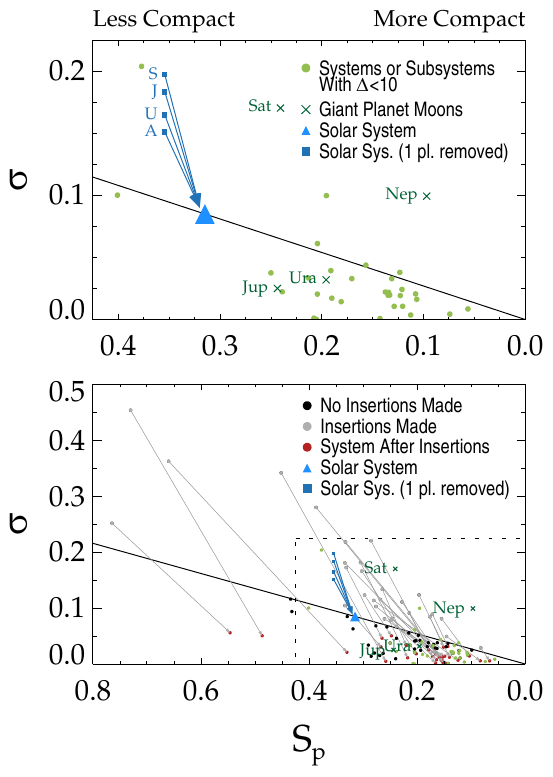}
\epsscale{1.0}
\caption{\small \emph{Top Panel:} $\sigma$ used in Equation~\ref{eq:chisquared} as a function of the average log period spacing between planets, $S_p$ (Equation~\ref{eq:compactness}), of multiple planet systems.
The thick line goes through two points: one point is the origin $(0,0)$. The other point ($S_p,\sigma$), is the sparseness of our Solar System (Equation \ref{eq:compactness}) and the $\sigma$
required for our Solar System to yield $\chi^2/\text{dof}=1$ in Equation~\ref{eq:chisquared}. 
The Solar System is shown as a blue triangle while the four small blue squares above it (from top to bottom) are the values of $(S_p, \sigma)$ from the Solar System if Saturn, Jupiter, Uranus and the Asteroid Belt are individually removed. 
The main prograde satellites of Jupiter, Saturn, Uranus and Neptune are also fit and indicated by small green crosses and labeled ``Jup'', ``Sat'', ``Ura'' and ``Nep'' respectively. 
The systems in our most complete sample (see Section~\ref{sec:mostcomplete}) are indicated by green dots.
\newline \emph{Bottom Panel:} Similar to the top panel but with the following exceptions. All systems are shown rather than only our most complete set. Black dots indicate systems where no insertion was made. 
Gray dots indicate a system where an insertion was made, and red dots indicate the $(S_p,\sigma)$ of the system after insertions have been made. Each gray point is connected to a red point by a gray arrow. The dashed box represents the range of the top panel, which encompasses all points in our most complete sample.}
\label{fig:compactness_sigma_TB2}
\end{figure}

In Figure~\ref{fig:compactness_sigma_TB2} we use the Solar System as a standard and make a linear fit between it and the (0,0) point. For each system in our analysis, the $\sigma$ in Equation~\ref{eq:chisquared} is calculated from this linear fit ($\sigma=0.270\,S_p$), given the sparseness/compactness of the system (Equation 6).
More compact systems are assigned smaller $\sigma$ values. 
Systems above the line adhere to the TB relation less well than the Solar System does, while those below the line adhere to the TB relation more closely than the Solar System does. 
Each time a hypothetical planet is inserted between detected planets, the system becomes more compact and a new, lower $\sigma$ is calculated. The periods of the inserted planets are calculated from the best fit TB relation. This ensures that the inserted planets make no contribution to the $\chi^2$ value.

\begin{figure}[p]
\plotone{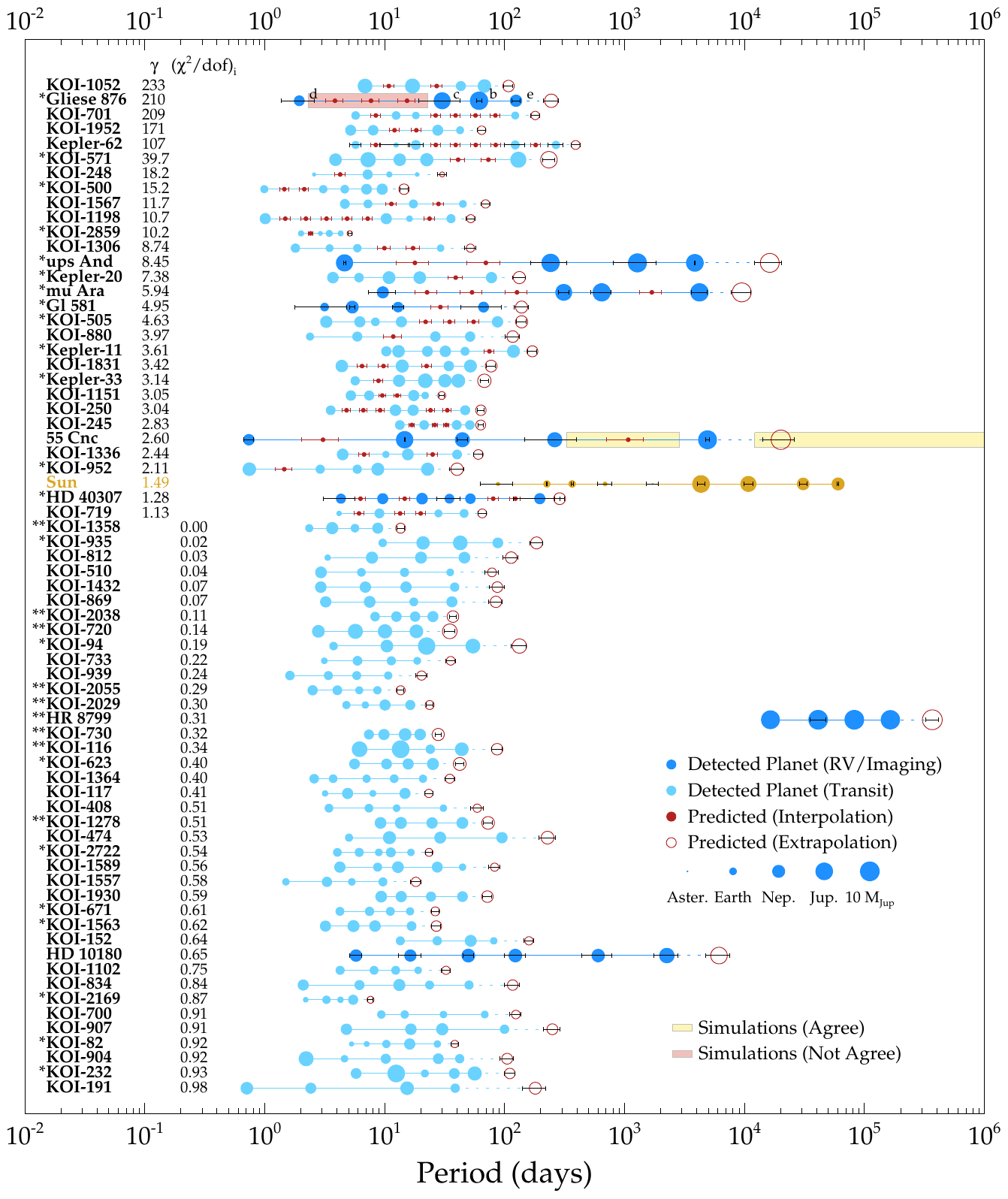}
\caption{\small Orbital periods of planets in multiple planet systems containing at least 4 detected planets. Systems where planet insertions have been made are sorted in descending order of the highest $\gamma$ found in each system (Equation~\ref{eq:gamma} and Table 2).
Systems without planet insertions (lower 2/3 of figure) are sorted in ascending order of $\chi^2/\text{dof}$ (Equation 4). 
Inserted and extrapolated planets are shown with red filled circles and red open circles respectively. The $\gamma$ value for the Solar System is calculated by excluding, then including, the Asteroid Belt. ``*'' indicates at least two adjacent planet pairs in the system have $\Delta$ values $<10$ if we had inserted a planet between each pair.
``**'' indicates all adjacent planet pairs have $\Delta$ values $<10$ if we had inserted an additional planet between each pair (see Section~\ref{sec:method}). 
For the systems where no planet was inserted by interpolation, predicted extrapolated planets are listed in Table 3. 
Planet letters are shown for Gliese 876.}
\label{fig:period_systems}
\end{figure}

\begin{figure}[p]
\plotone{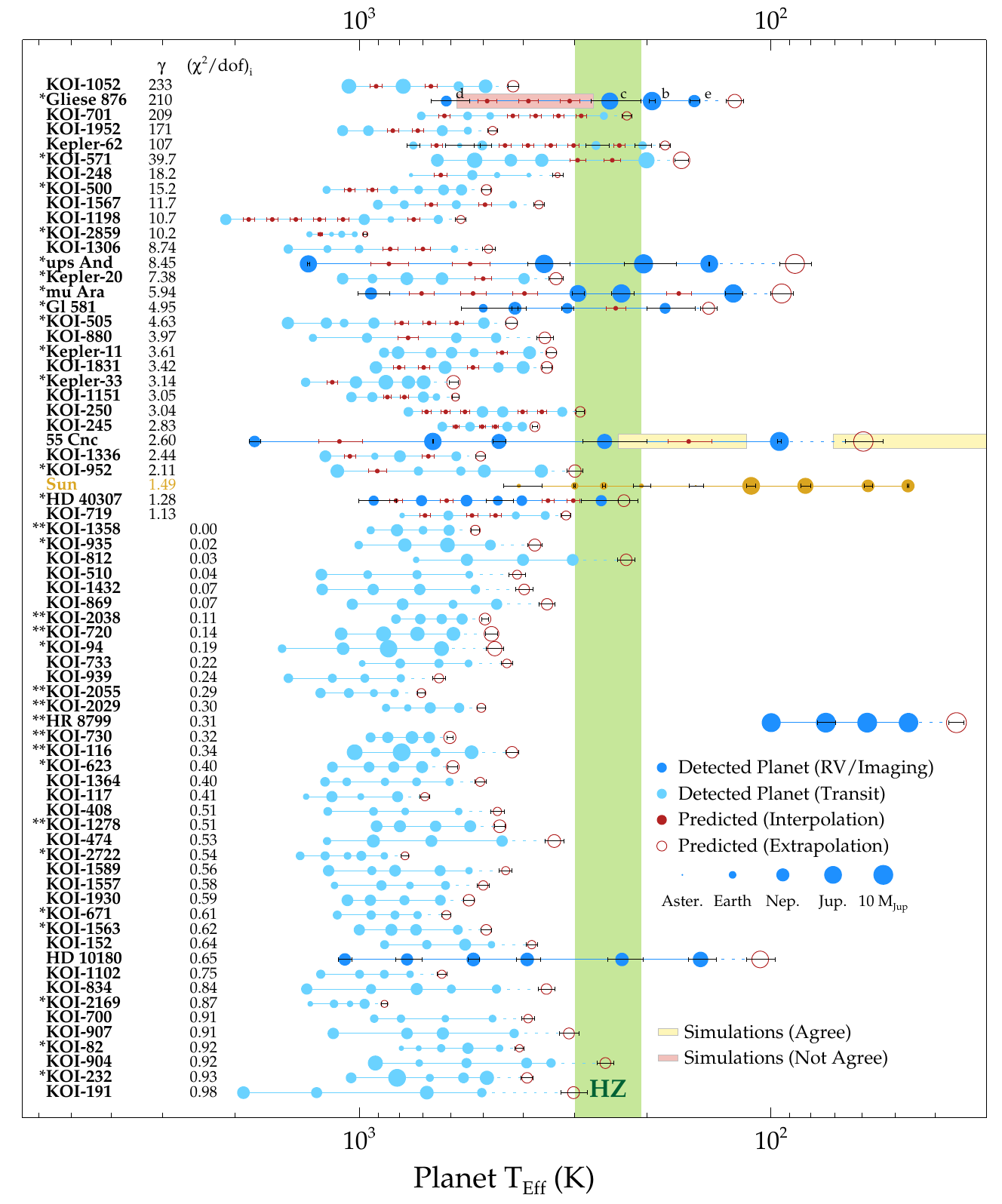}
\caption{\small Same as the previous figure except that the periods of the planets have been converted into effective planet temperatures (assuming an Earth-like albedo of 0.3), based on the luminosity of the stellar host.
The vertical green bar represents an approximate habitable zone using the effective temperature range between Venus and Mars, using an albedo of 0.3.
We estimate that the average number of planets in the habitable zone (HZ) of a star is approximately 1-2.
This estimate is based on i) the assumption that planetary systems extend across the HZ, and ii) on the number of 
HZ planets in the planetary systems shown here that span the HZ.}
\label{fig:Teff_systems}
\end{figure}

If a system has a $\chi^2/\text{dof}$ similar to or lower than the Solar System ($\lesssim 1$) before any insertions, no insertions are made. 
Applying our $\gamma$ method to our sample results in insertions being made in 29 out of 68 exosystems. For all systems in our sample we predict the location of the next outermost planet (extrapolation). Our planet predictions are shown in Figures~\ref{fig:period_systems} \& \ref{fig:Teff_systems} and Tables 2 \& 3.

\subsection{Upper Mass or Radius Limit}
\label{sec:constraints}
We can place constraints on the maximum mass (radial velocity detections) or maximum radius (transit detections) of our predicted planets. $M_\text{max}$ and $R_\text{max}$ are calculated by applying the lowest signal-to-noise ratio (SNR) of the detected planets in the same system, to the period of the inserted planet. That is, we calculate the maximum mass or radius of a planet at the predicted period that could avoid detection based on the lowest SNR of the detected planets in that same system. For transiting planets in the same system, SNR $\propto r^2 P^{-1/2}$ where $r$ is the planet radius and $P$ is the planet period. For radial velocity detected planets in the same system, SNR $\propto KP^{-5/6} \propto mP^{-7/6}$ where $K$ is the semi-amplitude of the radial velocity and $m$ is the planet mass. Maximum masses and radii for predicted planets can be found in Table 2.

\begin{threeparttable}
\newcommand{\tablesize}{\fontsize{5.5}{5.6}\selectfont}
\renewcommand{\arraystretch}{0.7}
\setlength{\tabcolsep}{3.0pt}
\tablesize

\caption{{\small Systems With Interpolated and Extrapolated Planet Predictions}}
\begin{tabular}{lccccccccccc}
\toprule[1pt]
\multirow{5}{*}{\bf{System}} & \multirow{5}{*}{\bf{\begin{tabular}{c} Number \\ Inserted \end{tabular}}} & \multirow{5}{*}{{\begin{tabular}{c} $\boldsymbol{\gamma}$ \\ (Eq. 5) \end{tabular}}} & \multirow{5}{*}{\bf{$\boldsymbol{\Delta\gamma}^{\,a}$}} & \multirow{5}{*}{\bf{$\boldsymbol{\left(\dfrac{\chi^2}{\text{dof}}\right)_i}$}} & \multirow{5}{*}{\bf{$\boldsymbol{\left(\dfrac{\chi^2}{\text{dof}}\right)_f}$}} & \multirow{5}{*}{\bf{\begin{tabular}{c} Inserted \\ Planet \# \end{tabular}}} & \multirow{5}{*}{\bf{\begin{tabular}{c} Period \\ (days) \end{tabular}}} & \multirow{5}{*}{\bf{\begin{tabular}{c} A \\ (AU) \end{tabular}}} & \multirow{5}{*}{\bf{\begin{tabular}{c} R$_\text{max}$$^{\,b}$ \\ (R$\boldsymbol{_\oplus}$) \end{tabular}}} & \multirow{5}{*}{\bf{\begin{tabular}{c} M$_\text{max}$$^{\,b}$ \\ (M$\boldsymbol{_\oplus}$) \end{tabular}}} & \multirow{5}{*}{\bf{\begin{tabular}{c} T$_\text{eff}$ \\ (K) \end{tabular}}}\\\\\\\\\\
\midrule
KOI-1052 & 2 & 234.0 & 1528.2 & 1.36 & 0.01 & 1 & $11\pm2$ & 0.10 & 1.6 & - & 909 \\
 &  &  &  &  &  & 2 & $28\pm3$ & 0.18 & 2.0 & - & 669 \\
 &  &  &  &  &  & \quad\;\;\,3 E$\,^\text{c}$ & $108\pm12$ & 0.46 & 2.8 & - & 423 \\
Gliese 876 & 3 & 210.7 & 32.5 & 7.88 & 0.02 & 1 & $3.9\pm0.7$ & 0.03 & - & 0.7 & 489 \\
 &  &  &  &  &  & 2 & $8\pm2$ & 0.05 & - & 1.2 & 388 \\
 &  &  &  &  &  & 3 & $16\pm3$ & 0.08 & - & 2.2 & 308 \\
 &  &  &  &  &  & \quad4 E & $245\pm40$ & 0.52 & - & 21.9 & 122 \\
KOI-701 & 5 & 209.2 & 58.2 & 5.92 & 0.01 & 1 & $8.5\pm0.8$ & 0.07 & 0.5 & - & 621 \\
 &  &  &  &  &  & 2 & $27\pm3$ & 0.15 & 0.7 & - & 423 \\
 &  &  &  &  &  & 3 & $39\pm4$ & 0.19 & 0.8 & - & 372 \\
 &  &  &  &  &  & 4 & $58\pm6$ & 0.25 & 0.9 & - & 328 \\
 &  &  &  &  &  & 5 & $84\pm8$ & 0.32 & 1.0 & - & \bf{288} \\
 &  &  &  &  &  & \quad6 E & $180\pm17$ & 0.54 & 1.2 & - & \bf{223} \\
KOI-1952 & 2 & 171.1 & 10.9 & 3.26 & 0.01 & 1 & $13\pm2$ & 0.10 & 1.5 & - & 828 \\
 &  &  &  &  &  & 2 & $19\pm2$ & 0.14 & 1.6 & - & 720 \\
 &  &  &  &  &  & \quad3 E & $65\pm7$ & 0.31 & 2.2 & - & 474 \\
Kepler-62 & 6 & 107.3 & 40.6 & 4.04 & 0.01 & 1 & $8.5\pm0.8$ & 0.07 & 0.5 & - & 649 \\
 &  &  &  &  &  & 2 & $27\pm3$ & 0.15 & 0.6 & - & 442 \\
 &  &  &  &  &  & 3 & $40\pm4$ & 0.20 & 0.7 & - & 389 \\
 &  &  &  &  &  & 4 & $58\pm6$ & 0.26 & 0.8 & - & 342 \\
 &  &  &  &  &  & 5 & $85\pm8$ & 0.33 & 0.9 & - & 301 \\
 &  &  &  &  &  & 6 & $182\pm18$ & 0.55 & 1.0 & - & \bf{233} \\
 &  &  &  &  &  & \quad7 E & $391\pm37$ & 0.92 & 1.2 & - & 180 \\
KOI-571 & 2 & 39.8 & 11.4 & 4.87 & 0.07 & 1 & $41\pm6$ & 0.19 & 0.8 & - & \bf{294} \\
 &  &  &  &  &  & 2 & $74\pm10$ & 0.28 & 1.0 & - & \bf{242} \\
 &  &  &  &  &  & \quad3 E & $234\pm32$ & 0.61 & 1.3 & - & 164 \\
KOI-248 & 1 & 18.3 & 17.5 & 2.48 & 0.13 & 1 & $4.3\pm0.5$ & 0.04 & 1.4 & - & 633 \\
 &  &  &  &  &  & \quad2 E & $31\pm4$ & 0.16 & 2.2 & - & 329 \\
KOI-500 & 2 & 15.2 & 4.6 & 5.42 & 0.18 & 1 & $1.5\pm0.2$ & 0.02 & 1.1 & - & 1055 \\
 &  &  &  &  &  & 2 & $2.2\pm0.2$ & 0.03 & 1.2 & - & 929 \\
 &  &  &  &  &  & \quad3 E & $15\pm2$ & 0.11 & 2.0 & - & 490 \\
KOI-1567 & 2 & 11.8 & 12.1 & 1.51 & 0.07 & 1 & $12\pm2$ & 0.10 & 2.0 & - & 668 \\
 &  &  &  &  &  & 2 & $29\pm3$ & 0.18 & 2.5 & - & 494 \\
 &  &  &  &  &  & \quad3 E & $70\pm8$ & 0.32 & 3.1 & - & 366 \\
KOI-1198 & 6 & 10.8 & 1.5 & 7.19 & 0.11 & 1 & $1.5\pm0.2$ & 0.03 & 1.2 & - & 1855 \\
 &  &  &  &  &  & 2 & $2.3\pm0.3$ & 0.04 & 1.3 & - & 1626 \\
 &  &  &  &  &  & 3 & $3.3\pm0.4$ & 0.05 & 1.4 & - & 1425 \\
 &  &  &  &  &  & 4 & $4.9\pm0.5$ & 0.06 & 1.6 & - & 1249 \\
 &  &  &  &  &  & 5 & $7.3\pm0.7$ & 0.08 & 1.7 & - & 1095 \\
 &  &  &  &  &  & 6 & $24\pm3$ & 0.17 & 2.3 & - & 737 \\
 &  &  &  &  &  & \quad7 E & $53\pm6$ & 0.29 & 2.8 & - & 566 \\
KOI-2859 & 1 & 10.2 & -1.0 & 1.69 & 0.16 & 1 & $2.41\pm0.10$ & 0.03 & 0.6 & - & 1242 \\
 &  &  &  &  &  & \quad2 E & $5.2\pm0.3$ & 0.05 & 0.8 & - & 967 \\
KOI-1306 & 2 & 8.8 & 0.6 & 4.14 & 0.23 & 1 & $10\pm2$ & 0.09 & 1.4 & - & 841 \\
 &  &  &  &  &  & 2 & $18\pm3$ & 0.13 & 1.6 & - & 700 \\
 &  &  &  &  &  & \quad3 E & $52\pm7$ & 0.27 & 2.1 & - & 485 \\
ups And & 2 & 8.5 & 1.0 & 5.31 & 0.30 & 1 & $18\pm6$ & 0.14 & - & 3.8 & 847 \\
 &  &  &  &  &  & 2 & $70\pm22$ & 0.36 & - & 11.9 & 537 \\
 &  &  &  &  &  & \quad3 E & $16300\pm5000$ & 13.57 & - & 1116.0 & 87 \\
Kepler-20 & 1 & 7.4 & 1.0 & 3.51 & 0.42 & 1 & $40\pm6$ & 0.22 & 1.2 & - & 500 \\
 &  &  &  &  &  & \quad2 E & $133\pm19$ & 0.49 & 1.6 & - & 332 \\
mu Ara & 4 & 6.0 & 1.8 & 4.15 & 0.17 & 1 & $23\pm5$ & 0.16 & - & 7.7 & 705 \\
 &  &  &  &  &  & 2 & $54\pm11$ & 0.29 & - & 15.8 & 529 \\
 &  &  &  &  &  & 3 & $127\pm26$ & 0.52 & - & 32.4 & 396 \\
 &  &  &  &  &  & 4 & $1690\pm350$ & 2.90 & - & 280.2 & 167 \\
 &  &  &  &  &  & \quad5 E & $9500\pm2000$ & 9.17 & - & 1180.7 & 94 \\
Gl 581 & 1 & 5.0 & 0.9 & 3.72 & 0.63 & 1 & $30\pm5$ & 0.13 & - & 3.0 & \bf{238} \\
 &  &  &  &  &  & \quad2 E & $139\pm23$ & 0.35 & - & 11.2 & 141 \\
KOI-505 & 3 & 4.7 & 0.6 & 8.20 & 0.56 & 1 & $22\pm3$ & 0.15 & 4.6 & - & 788 \\
 &  &  &  &  &  & 2 & $35\pm4$ & 0.20 & 5.1 & - & 676 \\
 &  &  &  &  &  & 3 & $56\pm7$ & 0.27 & 5.8 & - & 580 \\
 &  &  &  &  &  & \quad4 E & $139\pm16$ & 0.51 & 7.3 & - & 426 \\
KOI-880 & 1 & 4.0 & 2.4 & 1.35 & 0.28 & 1 & $12\pm2$ & 0.10 & 2.1 & - & 761 \\
 &  &  &  &  &  & \quad2 E & $117\pm20$ & 0.45 & 3.7 & - & 354 \\
Kepler-11 & 1 & 3.7 & 1.7 & 3.15 & 0.69 & 1 & $75\pm8$ & 0.34 & 2.9 & - & 449 \\
 &  &  &  &  &  & \quad2 E & $171\pm17$ & 0.59 & 3.5 & - & 341 \\
KOI-1831 & 3 & 3.5 & 1.7 & 2.54 & 0.23 & 1 & $6.5\pm0.7$ & 0.06 & 0.9 & - & 800 \\
 &  &  &  &  &  & 2 & $9.8\pm1.0$ & 0.08 & 1.0 & - & 697 \\
 &  &  &  &  &  & 3 & $23\pm3$ & 0.15 & 1.2 & - & 529 \\
 &  &  &  &  &  & \quad4 E & $78\pm8$ & 0.34 & 1.6 & - & 350 \\
Kepler-33 & 1 & 3.2 & 0.1 & 3.69 & 0.90 & 1 & $8.9\pm0.8$ & 0.09 & 1.9 & - & 1162 \\
 &  &  &  &  &  & \quad2 E & $68\pm7$ & 0.35 & 3.2 & - & 590 \\
KOI-1151 & 2 & 3.1 & 0.2 & 3.77 & 0.54 & 1 & $9.6\pm0.7$ & 0.09 & 0.8 & - & 854 \\
 &  &  &  &  &  & 2 & $12.7\pm0.9$ & 0.10 & 0.9 & - & 776 \\
 &  &  &  &  &  & \quad3 E & $30\pm2$ & 0.18 & 1.1 & - & 584 \\
KOI-250 & 5 & 3.1 & 2.5 & 2.26 & 0.14 & 1 & $4.9\pm0.4$ & 0.05 & 1.2 & - & 686 \\
 &  &  &  &  &  & 2 & $6.7\pm0.6$ & 0.06 & 1.3 & - & 616 \\
 &  &  &  &  &  & 3 & $9.2\pm0.8$ & 0.07 & 1.4 & - & 553 \\
 &  &  &  &  &  & 4 & $25\pm2$ & 0.14 & 1.8 & - & 401 \\
 &  &  &  &  &  & 5 & $34\pm3$ & 0.17 & 2.0 & - & 360 \\
 &  &  &  &  &  & \quad6 E & $64\pm5$ & 0.27 & 2.3 & - & \bf{290} \\
KOI-245 & 3 & 2.9 & 2.4 & 1.56 & 0.17 & 1 & $16.8\pm0.9$ & 0.12 & 0.3 & - & 582 \\
 &  &  &  &  &  & 2 & $27\pm2$ & 0.16 & 0.4 & - & 502 \\
 &  &  &  &  &  & 3 & $33\pm2$ & 0.18 & 0.4 & - & 467 \\
 &  &  &  &  &  & \quad4 E & $64\pm4$ & 0.28 & 0.5 & - & 374 \\
55 Cnc & 2 & 2.6 & 0.5 & 1.49 & 0.25 & 1 & $4\pm2$ & 0.04 & - & 1.4 & 1117 \\
 &  &  &  &  &  & 2 & $1080\pm370$ & 1.98 & - & 178.2 & 158 \\
 &  &  &  &  &  & \quad3 E & $20100\pm6900$ & 13.97 & - & 2046.5 & 59 \\
KOI-1336 & 2 & 2.5 & 11.0 & 1.07 & 0.19 & 1 & $6.8\pm0.7$ & 0.07 & 1.7 & - & 1053 \\
 &  &  &  &  &  & 2 & $26\pm3$ & 0.17 & 2.4 & - & 679 \\
 &  &  &  &  &  & \quad3 E & $61\pm6$ & 0.31 & 3.0 & - & 507 \\
KOI-952 & 1 & 2.2 & 0.1 & 2.36 & 0.76 & 1 & $1.5\pm0.3$ & 0.02 & 0.9 & - & 904 \\
 &  &  &  &  &  & \quad2 E & $41\pm7$ & 0.19 & 2.1 & - & \bf{299} \\
HD 40307 & 4 & 1.3 & 0.2 & 1.91 & 0.32 & 1 & $6.3\pm0.7$ & 0.06 & - & 0.4 & 814 \\
 &  &  &  &  &  & 2 & $15\pm2$ & 0.11 & - & 0.8 & 613 \\
 &  &  &  &  &  & 3 & $81\pm9$ & 0.33 & - & 3.4 & 348 \\
 &  &  &  &  &  & 4 & $123\pm13$ & 0.44 & - & 4.8 & 302 \\
 &  &  &  &  &  & \quad5 E & $287\pm30$ & 0.78 & - & 9.7 & \bf{227} \\
KOI-719 & 3 & 1.2 & 0.1 & 1.32 & 0.31 & 1 & $6.2\pm0.6$ & 0.06 & 0.6 & - & 692 \\
 &  &  &  &  &  & 2 & $14\pm2$ & 0.10 & 0.7 & - & 532 \\
 &  &  &  &  &  & 3 & $20\pm2$ & 0.13 & 0.8 & - & 466 \\
 &  &  &  &  &  & \quad4 E & $66\pm7$ & 0.28 & 1.1 & - & 314 \\
\midrule
\midrule[1pt]

\end{tabular}
\begin{tablenotes}
\scriptsize{
\item[a]{$\Delta\gamma=(\gamma_1-\gamma_2)/\gamma_2$ where $\gamma_1$ and $\gamma_2$ are the highest and second highest $\gamma$ values for that system respectively.}
\item[b]{$R_\text{max}$ and $M_\text{max}$ are calculated by applying the lowest SNR of the detected planets in the system to the period of the inserted planet. i.e. SNR $\propto r^2 P^{-1/2}$ for the planetary radius, where $r$ is the planet radius and $P$ is the planet period. SNR $\propto KP^{-5/6}$ for planets detected by radial velocity, where $K$ is the semi-amplitude velocity (see Section 4.3).}
\item[c]{A planet number followed by ``E" indicates the planet is extrapolated (has a larger period than the outermost detected planet in the system).}}
\end{tablenotes}
\end{threeparttable}

\begin{threeparttable}
\newcommand{\tablesize}{\fontsize{7.8}{7.9}\selectfont}
\renewcommand{\arraystretch}{0.95}
\setlength{\tabcolsep}{5.0pt}
\tablesize
\caption{{\small Systems With Only Extrapolated Planet Predictions}}
\begin{tabular}{lccccccc}

\toprule[1pt]
\multirow{3}{*}{\bf{System}} & \multirow{3}{*}{\bf{$\boldsymbol{\left(\dfrac{\chi^2}{\text{dof}}\right)_i}$}} & \multirow{3}{*}{\bf{\begin{tabular}{c} Period \\ (days) \end{tabular}}} & \multirow{3}{*}{\bf{\begin{tabular}{c} A \\ (AU) \end{tabular}}} & \multirow{3}{*}{\bf{\begin{tabular}{c} R$_\text{max}$$^{\,a}$ \\ (R$\boldsymbol{_\oplus}$) \end{tabular}}} & \multirow{3}{*}{\bf{\begin{tabular}{c} M$_\text{max}$$^{\,a}$ \\ (M$\boldsymbol{_\oplus}$) \end{tabular}}} & \multirow{3}{*}{\bf{\begin{tabular}{c} T$_\text{eff}$ \\ (K) \end{tabular}}}\\\\\\
\midrule
KOI-1358 & 0.01 & $13.6\pm1.0$ & 0.10 & 1.6 & - & 522 \\
KOI-935 & 0.03 & $185\pm23$ & 0.68 & 4.1 & - & 374 \\
KOI-812 & 0.04 & $114\pm17$ & 0.38 & 2.5 & - & \bf{224} \\
KOI-510 & 0.05 & $79\pm11$ & 0.34 & 3.5 & - & 413 \\
KOI-1432 & 0.07 & $87\pm13$ & 0.38 & 2.0 & - & 397 \\
KOI-869 & 0.08 & $85\pm12$ & 0.35 & 3.8 & - & 349 \\
KOI-2038 & 0.11 & $38\pm3$ & 0.21 & 1.8 & - & 494 \\
KOI-720 & 0.14 & $35\pm4$ & 0.20 & 2.9 & - & 477 \\
KOI-94 & 0.19 & $133\pm20$ & 0.55 & 3.5 & - & 468 \\
KOI-733 & 0.22 & $36\pm4$ & 0.20 & 2.8 & - & 437 \\
KOI-939 & 0.24 & $21\pm3$ & 0.15 & 1.9 & - & 640 \\
KOI-2055 & 0.29 & $13.6\pm1.0$ & 0.11 & 1.3 & - & 706 \\
HR 8799 & 0.31 & $368000\pm46000$ & 115.30 & - & 3730.5 & 35 \\
KOI-2029 & 0.31 & $24\pm2$ & 0.15 & 0.9 & - & 505 \\
KOI-730 & 0.32 & $28\pm2$ & 0.18 & 2.8 & - & 602 \\
KOI-116 & 0.34 & $87\pm10$ & 0.38 & 1.3 & - & 425 \\
KOI-623 & 0.40 & $43\pm4$ & 0.23 & 1.3 & - & 592 \\
KOI-1364 & 0.41 & $35\pm4$ & 0.20 & 3.0 & - & 508 \\
KOI-117 & 0.42 & $24\pm2$ & 0.17 & 1.4 & - & 692 \\
KOI-408 & 0.51 & $60\pm7$ & 0.29 & 2.6 & - & 461 \\
KOI-1278 & 0.52 & $73\pm7$ & 0.35 & 2.0 & - & 455 \\
KOI-474 & 0.54 & $228\pm37$ & 0.77 & 3.8 & - & 336 \\
KOI-2722 & 0.54 & $24\pm2$ & 0.17 & 1.3 & - & 774 \\
KOI-1589 & 0.56 & $83\pm9$ & 0.37 & 2.4 & - & 440 \\
KOI-1557 & 0.58 & $19\pm2$ & 0.12 & 2.0 & - & 499 \\
KOI-1930 & 0.60 & $72\pm7$ & 0.35 & 2.3 & - & 541 \\
KOI-1563 & 0.62 & $27\pm3$ & 0.17 & 3.7 & - & 491 \\
KOI-671 & 0.62 & $27\pm2$ & 0.17 & 1.5 & - & 614 \\
KOI-152 & 0.64 & $160\pm16$ & 0.60 & 5.4 & - & 381 \\
HD 10180 & 0.65 & $6200\pm1500$ & 6.67 & - & 147.9 & 106 \\
KOI-1102 & 0.75 & $33\pm3$ & 0.20 & 2.8 & - & 630 \\
KOI-834 & 0.84 & $117\pm17$ & 0.47 & 2.5 & - & 351 \\
KOI-2169 & 0.87 & $7.7\pm0.4$ & 0.07 & 0.7 & - & 868 \\
KOI-700 & 0.91 & $124\pm14$ & 0.48 & 2.1 & - & 388 \\
KOI-907 & 0.91 & $250\pm41$ & 0.76 & 4.5 & - & 309 \\
KOI-904 & 0.92 & $105\pm14$ & 0.37 & 2.2 & - & \bf{252} \\
KOI-82 & 0.92 & $39\pm3$ & 0.21 & 0.9 & - & 408 \\
KOI-232 & 0.93 & $111\pm12$ & 0.46 & 2.1 & - & 391 \\
KOI-191 & 0.99 & $180\pm39$ & 0.61 & 3.4 & - & 301 \\
\midrule
\midrule[1pt]

\end{tabular}
\begin{tablenotes}
\item[a]{$R_\text{max}$ and $M_\text{max}$ are calculated by applying the lowest SNR of the detected planets in the system to the period of the inserted planet. i.e. SNR $\propto r^2 P^{-1/2}$ for the planetary radius, where $r$ is the planet radius and $P$ is the planet period. SNR $\propto KP^{-5/6}$ for planets detected by radial velocity, where $K$ is the semi-amplitude velocity (see Section 4.3).}
\end{tablenotes}
\end{threeparttable}

\section{DISCUSSION}
\label{sec:discussion}

As exoplanet detection sensitivities improve, the number of multiple-planet systems will increase, along with the average number of planets detected in those system. Finding additional planets in these systems can be aided by making constrained predictions about their locations. This helps both in the observation stage where the sampling can be optimized to a specific period and in the analysis stage where more care can be taken at predicted locations.

Some planet predictions have already been made by numerical integration. Regions are mapped in semi-major axis and eccentricity space where additional planets can be inserted while the system remains stable. Due to computing constraints, numerical integrations tend to focus on two-planet systems. For example, in our sample of systems containing four or more planets, only two systems have been analyzed for the stability of additional planets by numerical integration. As the number of planets detected per system and the number of systems continues to increase, directly integrating each system becomes more impractical. We emphasize that we do not expect the generalized TB relation can be applied to every type system (e.g. where exact resonances are present).

\subsection{Comparison With TB Relation Studies}

\indent\emph{55 Cnc:}\\
\cite{Poveda2008} represented the semi-major axes of the planets in 55 Cnc with an exponential function (Equation~\ref{eq:poveda}). They claimed a good fit when the 5th planet in the system was assigned $n=6$, leaving a gap at $n=5$ for an additional planet beyond the detection threshold. The fit was extrapolated to $n=7$ to predict a planet beyond the outermost detected planet. The semi-major axes of the predicted planets were $\sim 2\text{ AU}$ (P=1,086 days) and $\sim 15\text{ AU}$ (P=22,309 days), with no attempt made to estimate uncertainties. After their paper was published, the period of the innermost planet was updated from 2.8 days to 0.7 days \citep{Winn2011a}.

When using the old value for the period, we similarly predict a single planet located at $2.0\pm0.5$ AU (P=1,111$\pm$406 days). When using the new value for the period, we predict two planets, the first at $0.04\pm0.01$ AU (P=3.1$\pm$1.1 days), while the prediction for the planet at $2.0\pm0.5$ AU (P=1,111$\pm$406 days) is unchanged. We calculate the location of a planet exterior to the outermost detected planet at $15.1\pm3.5$ AU (P=22,988$\pm$7,816 days) and $14.7\pm3.4$ AU (P=22,079$\pm$7,492 days) when the old and new data are used respectively. All predictions are compatible with those made by \cite{Poveda2008}, with the exception of our additional planet prediction at $0.04\pm0.01$ AU (P=3.1$\pm$1.1 days) caused by the updated period of the innermost planet.

\cite{Cuntz2011} applied the 3 parameter TB relation (Equation~\ref{eq:appendix3TB}) to 55 Cnc. The residuals were minimized when $A\simeq0.037$, $B\simeq0.048$ and $Z\simeq1.98$ and for planet numbers $n=-\infty$, 1, 2, 4 and 7. This implies undetected planets occupying the $n=0$, 3, 5 and 6 spaces, calculated to have semi-major axes of 0.085 (P=9.5 days), 0.41 (P=100 days), 1.50 (P=705 days) and 2.95 AU (P=1,945 days). Cuntz predicted a planet exterior to the outermost planet ($n=8$) located at $11.8\pm0.9$ AU (P=15,599$\pm$1,780 days). The predicted planets at $n=3$ and $n=8$ are within the uncertainties of our predictions. The remaining three predicted planets are incompatible with our predictions, although no uncertainty estimate is made by Cuntz.

\subsection{Comparison With Numerical Simulations}

\indent\emph{55 Cnc:}\\
Our prediction of a planet at $2.0\pm0.5$ AU (P=1,111$\pm$406 days) is compatible with numerical simulations which have found that the 55 Cnc system can support an additional planet between 0.9 (P=327 days) and 3.8 AU (P=2,844 days) with an eccentricity below 0.4 \citep{Raymond2008,Ji2009}. \cite{Raymond2008} also demonstrate that a planet exterior to the outermost detected planet will likely reside beyond 10 AU (P=12,143 days), compatible with our more explicit prediction of $14.7\pm3.4$ AU (P=22,079$\pm$7,492 days).

\emph{Gl 876:}\\
Numerical simulations by \cite{Correia2010} show a small stable region for an additional planet centered on 0.08 AU (P=14.3 days) around the M dwarf Gliese 876. This region is stable due to the $2:1$ mean motion resonance with Gliese 876 c, the $\sim0.7$ Jupiter mass planet located at 0.13 AU (P=30 days). \cite{Gerlach2012} initially found a similar stable region around $0.08$ AU (P=14 days). Further analysis suggested that initially stable orbits became chaotic and unstable over larger time frames. In our analysis, we predict 3 additional planets between Gliese 876 c and the innermost planet, Gliese 876 d. The inserted planets have semi-major axes ranging from 0.03 (P=3.2 days) to 0.1 AU (P=20 days). Although the numerical simulations of \cite{Gerlach2012} suggest that these planets are located within an unstable region, our predictions correlate to their most stable areas within that zone.

\subsection{Potential Habitable Zone Predictions}
\emph{Extrapolated Systems}:
For systems which adhered to the TB relation well without any insertions, we predicted the period of the next outermost planet in the system.
This resulted in a prediction of a potentially habitable terrestrial planet in three Kepler candidate systems. KOI-812 is a four-planet system with $\chi^2/\text{dof}=0.04$ (compared to 1.0 for the Solar System). 
When we insert a 5th planet into the system, it has an estimated effective temperature (assuming an Earth-like albedo of 0.3) of $\sim224$ K and a maximum radius of $2.5$ R$_\oplus$ based on its current non-detection. 
KOI-904 is a five-planet Kepler candidate system where we have predicted an extrapolated 6th planet. In this case, we predict an undetected planet with an effective temperature of $\sim252$ K and a maximum radius of $2.2$ R$_\oplus$.

\subsection{Kepler Candidates With The Highest Probability Of Validation}
\label{sec:acrit}
Assuming a high degree of coplanarity \citep{Fang2012a}, and using the angular-momentum-weighted average inclination of planets in each system, \textless$i$\textgreater$_L$, we compute a critical semi-major axis $a_\text{crit}$. Beyond this value, a coplanar planet would no longer transit (impact parameter $b > 1$, where $b=a\cos$\textless$i$\textgreater$_L/R_*$ and $R_*$ is the estimated radius of the host star). We compute these $a_\text{crit}$ values for each transiting system in our sample. According to the reported Kepler data\footnote{http://exoplanetarchive.ipac.caltech.edu} KOI-1102, KOI-623, KOI-719, KOI-720 and KOI-939 are the Kepler candidate systems where an undetected planet exterior to each outermost detected planet is most likely to transit i.e., the semi-major axis $a_\text{out}$ of the outermost detected planet contributes to a low value of $a_\text{out}/a_\text{crit}$. 


\subsection{The Two and Three Parameter TB Relations}
\label{sec:2v3param}
To check the robustness of our 2 parameter TB relation predictions, we repeated our analysis using the 3 parameter TB relation (Equation~\ref{eq:appendix3TB}). 
The two forms of the TB relation do not yield identical results.
For example, when the 3 parameter TB relation is applied to the Gliese 876 system, no planet insertions are made. This is a result of the decoupling of the distance to the first planet from the rest of the system (Figure~\ref{fig:illustration}, Appendix~\ref{sec:2v3paramapp}). 
The 2nd, 3rd and 4th planets in Gliese 876 fit the 3 parameter TB relation well while there is a large gap between the 1st and 2nd planets. 
However, this does not affect the fit since the distance to the first planet is its own decoupled parameter. 
This is not unique to the Gliese 876 system and the 3 parameter TB relation often does not insert a planet between the first and second detected planets, even if a large gap is present. 
Of the 29 systems where we predicted interpolated planets, 18 of these systems are also predicted to have interpolated planets by the 3 parameter TB relation. In 8 of the 11 systems where planets are inserted by the 2 parameter TB relation but not by the 3 parameter TB relation, there is a significant gap between the first and second planet in each system. 
The prioritization of the lists of predicted planets by the 2 and 3 parameter TB fits also varies due to the different $\gamma$ values produced for a given system by each form of the TB relation.

\section{CONCLUSION}
\label{sec:conclusion}

We first identified a sample of exoplanet systems most likely to be complete based on the dynamical spacing $\Delta$ of planet pairs. We applied the TB relation to each system in the sample and found that $\sim94\%$ of the systems adhered to the TB relation to approximately the same extent or better than the Solar System. This was taken as evidence that the TB relation can be applied to exoplanet systems in general. 

We then applied the TB relation to the systems most likely to be incompletely sampled and we made planet predictions in these systems. Our method involved inserting up to 10 hypothetical planets into each system and we chose the insertion solution which maximized $\gamma$ (Equation~\ref{eq:gamma}), which is a measure of the improvement in the $\chi^2/\text{dof}$ value (Equation~\ref{eq:chisquared}) per planet inserted. We made planet predictions (interpolations) inserting 73 planets into 29 of the 68 systems analyzed here. For all 68 systems we predicted the period of the next outermost planet (extrapolation).

The maximum mass or radius for each predicted planet was calculated based on the detection limit in the host system.
Since the TB relation may be mass-independent (e.g. the Asteroid Belt), our predicted planets do not have a lower mass limit. 
Thus, some significant fraction of our predictions probably correspond to low mass planets or Asteroid Belt analogues, in which case, validation in the near future is unlikely. Our predictions include a number of potentially habitable terrestrial planets. We identified Kepler candidate systems where an undetected planet exterior to the outermost detected planet has the greatest chance of transiting, assuming near coplanarity.

As the number of candidate planets in a system increases, the system's light curve or radial velocity curve becomes more complicated. This increases the difficulty of detecting additional planets within the system. 
Using the TB relation to predict planetary orbital periods (Tables 2 \& 3) narrows the region of parameter space to be searched, and increases the likelihood of detecting an additional planet.
We expect our predictions to be partially tested using new data from the Kepler mission in the near future.
If a significant number of our predictions are substantiated, we expect the TB relation to become an important detection tool for systems which already contain a number of planets.

\section{ACKNOWLEDGEMENTS}
This research has made use of the NASA Exoplanet Archive, which is operated by the California Institute of Technology, under contract with the National Aeronautics and Space Administration under the Exoplanet Exploration Program. This research has made use of the Exoplanet Orbit Database
and the Exoplanet Data Explorer at exoplanets.org. We thank Steffen Jacobsen for developing the procedure for calculating the $a_\text{crit}$ values used in Section~\ref{sec:acrit}.

\appendix
\section{THE TWO AND THREE PARAMETER TB RELATIONS}
\label{sec:2v3paramapp}
The original TB relation was given a mathematical form by \cite{Wurm1787},
\begin{equation}
a_n=A+BZ^n, \hspace{10pt}n=-\infty,0,1,2,...
\label{eq:appendix3TB}
\end{equation}
where $a_n$ is the semi-major axis of the planet. The index $n=-\infty$ is assigned to the first planet, $n=0$ to the second planet, $n=1$ to the third planet and so on. Historically, $A=0.4$, $B=0.3$ and $Z=2.0$ for the Solar System, while the best fit values are $A\simeq0.382$, $B\simeq0.334$, $Z\simeq1.925$. 
The two parameter TB relation can be written as
\begin{equation}
a_n=aC^n, \hspace{10pt} n=1,2,3,...
\label{eq:appendix2TB}
\end{equation}
The exponential function given by \cite{Poveda2008} is equivalent to Eq.~\ref{eq:appendix2TB}:
\begin{equation}
\label{eq:poveda}
a_n=ae^{\lambda n},\hspace{8pt}=aC^n n=1,2,3,...\hspace{8pt}(C=e^\lambda)
\end{equation}
By comparing Equation~\ref{eq:appendix3TB} and Equation~\ref{eq:appendix2TB} we can see they are most similar for large $n$. Therefore the extra term in Equation~\ref{eq:appendix3TB} and the numbering methodology $n=-\infty,0,1,...$ produces a difference in the fits to the inner part of a planetary system.
This has led to the location of the Earth to be called peculiar \citep{Neslusan2004}.

Other analyzes of the Solar System tend to use the generalized 2 parameter TB relation as we have done in this paper \citep{Prentice1978,Basano1979,Neuhauser1986,Zhong-Wei1987,Graner1994}.

\begin{figure}[htb]
\centering
\subfloat[The 2 parameter TB relation (Equation~\ref{eq:appendix2TB})]{\includegraphics[width=0.47\textwidth]{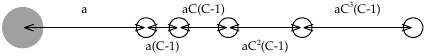}}
\;\;\;\;\;\;\;
\subfloat[The 3 parameter TB relation (Equation~\ref{eq:appendix3TB})]{\includegraphics[width=0.47\textwidth]{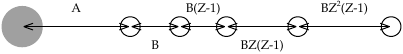}}
\caption{\small Illustration of the parameters in the two forms of the Titius-Bode relation (Equations~\ref{eq:appendix3TB} and~\ref{eq:appendix2TB}). Labels show the distances between adjacent objects. The large filled circle represents the star while the smaller open circles represent the planets.}
\label{fig:illustration}
\end{figure}

\section{THE DYNAMICAL SPACING $\Delta$}
\label{sec:delta}
The dynamical spacing $\Delta$ \citep{Gladman1993,Chambers1996} between two planets in the same planetary system with semi-major axes $a_1$ and $a_2$ is,
\begin{equation}
\Delta=\dfrac{a_2-a_1}{R_H}
\label{eq:dynamicalspacing}
\end{equation}
where $R_H$ is the two planets' mutual Hill radius, given by
\begin{equation}
R_H=\dfrac{a_1+a_2}{2}\left(\dfrac{m_1+m_2}{3M_*}\right)^{1/3}
\label{eq:mutualhillrad}
\end{equation}
where $m_1$ and $m_2$ are the masses the inner and outer planets respectively. $M_*$ is the mass of the host star.

{\scriptsize
\bibliographystyle{apj_withtitles}
\bibliography{planetspacing}
}

\end{document}